\documentclass[twocolumn,floats,floatfix,prl,superscriptaddress,showpacs,aps]{revtex4-1}
\usepackage{graphicx}%
\usepackage{dcolumn}%
\usepackage{bm}%
\begin{document}

\preprint{APS/123-QED}

\title{Metal oxide resistive switching: evolution of the density of states across the metal-insulator transition}

\author{A. Mottaghizadeh}
\affiliation{Laboratoire de Physique et d'Etude des Mat\'eriaux, UMR 8213, ESPCI-ParisTech-CNRS-UPMC, 10 rue Vauquelin, 75231 Paris, France}
\author{Q. Yu}
\affiliation{Laboratoire de Physique et d'Etude des Mat\'eriaux, UMR 8213, ESPCI-ParisTech-CNRS-UPMC, 10 rue Vauquelin, 75231 Paris, France}
\author{P.L. Lang}
\affiliation{Laboratoire de Physique et d'Etude des Mat\'eriaux, UMR 8213, ESPCI-ParisTech-CNRS-UPMC, 10 rue Vauquelin, 75231 Paris, France}
\author{A. Zimmers}
\email{Alexandre.Zimmers@espci.fr} \affiliation{Laboratoire de Physique et d'Etude des Mat\'eriaux, UMR 8213, ESPCI-ParisTech-CNRS-UPMC, 10 rue Vauquelin, 75231 Paris, France}
\author{H. Aubin}
\email{Herve.Aubin@espci.fr} \affiliation{Laboratoire de Physique et d'Etude des Mat\'eriaux, UMR 8213, ESPCI-ParisTech-CNRS-UPMC, 10 rue Vauquelin, 75231 Paris, France}

\date{\today}

\begin{abstract}
We report the study of metal-STO-metal memristors where the doping concentration in STO can be fine-tuned through electric field migration of oxygen vacancies. In this tunnel junction device, the evolution of the Density Of States (DoS) can be followed continuously across the Metal-Insulator Transition (MIT). At very low dopant concentration, the junction displays characteristic signatures of discrete dopants levels. As the dopant concentration increases, the semiconductor band gap fills in but a soft Coulomb gap remains. At even higher doping, a transition to a metallic state occurs where the DoS at the Fermi level becomes finite and Altshuler-Aronov corrections to the DoS are observed. At the critical point of the MIT, the DoS scales linearly with energy $N(\varepsilon) \sim \varepsilon$, the possible signature of multifractality.

\end{abstract}

% PACS,

\maketitle

Memristive devices have attracted considerable attention since the recognition that two-terminals resistive switching devices~\cite{Strukov2008} represent an example of a memristive element, whose existence was hypothesized by L. Chua in 1971. Numerous works on binary metal oxides MO$_x$ and STO, see Refs.~\cite{Dearnaley1970,Yang2013,Waser2007} for review, have demonstrated that resistance switching is due to the electric-field induced migration of positively charged oxygen vacancies. This anion migration leads to a change in the metal cation valence and therefore to a change in the band occupation~\cite{Yang2008}. In titanium based oxide materials, the oxidation number of the titanium cation changes from +IV to +III which corresponds to n-type doping. Doping dependency studies of semiconductors have been essential for the understanding of the Anderson localisation and the interplay between Coulomb interactions and disorder in the vicinity of MITs~\cite{Belitz1994,Lee1985}. While past works usually relied on the study of samples with fixed doping levels~\cite{Lee2004a,Bielejec2001,Lee1999a,Massey1996,McMillan1981}, we found that electric field induced migration of oxygen vacancies at very low temperature allows fine tuning of the dopants concentration from the insulating to the metallic regime. This makes resistive switching devices a remarkable tool for the study of MITs.

%These recent developments may lead to the emergence of a new branch of electronics where the manipulation of ions in electronic devices will be used to control their internal state to turn an electron conducting channel on or off, to store a memory element or, possibly, to change the electronic order within the device from an insulator to a conductor or even a superconductor.

We present a study of MIT in STO through electric field induced migration of oxygen vacancies at very low temperature ($T\sim 260~mK$). STO is a material with good oxygen mobility whose resistive switching properties have been much studied~\cite{Yang2008,Andreasson2009,Waser2009,Andreasson2007,Janousch2007,Szot2006,Meijer2005} where it has been demonstrated that nano-sized conducting filaments down to a single dislocation could be obtained~\cite{Szot2006}.
On a (110) oriented STO substrate, two electrodes of width $500~nm$ separated by a distance of $200~nm$ are deposited, Fig. 1a. A conducting filament is formed at room temperature, as illustrated Fig. 1a, by ramping up the voltage until a resistance switching is observed, Fig. 1c. This is the so-called forming step where oxygen vacancies accumulate along some percolating path between the two electrodes. This leads to a junction of resistance of only a few hundred Ohms, which is nearly temperature independent down to low temperature where it starts to decrease below $T\sim350~mK$, Fig. 1d, indicating the apparition of superconductivity. Indeed, doped STO is known to become superconducting in this temperature range\cite{Koonce1967}, with $T_c \sim 350 ~mK$ for a carrier concentration of the order of $10^{20}~e~cm^{-3}$, Ref.~\cite{Ueno2011}.

\begin{figure}[h!]
\begin{center}
\includegraphics[width=8cm,keepaspectratio]{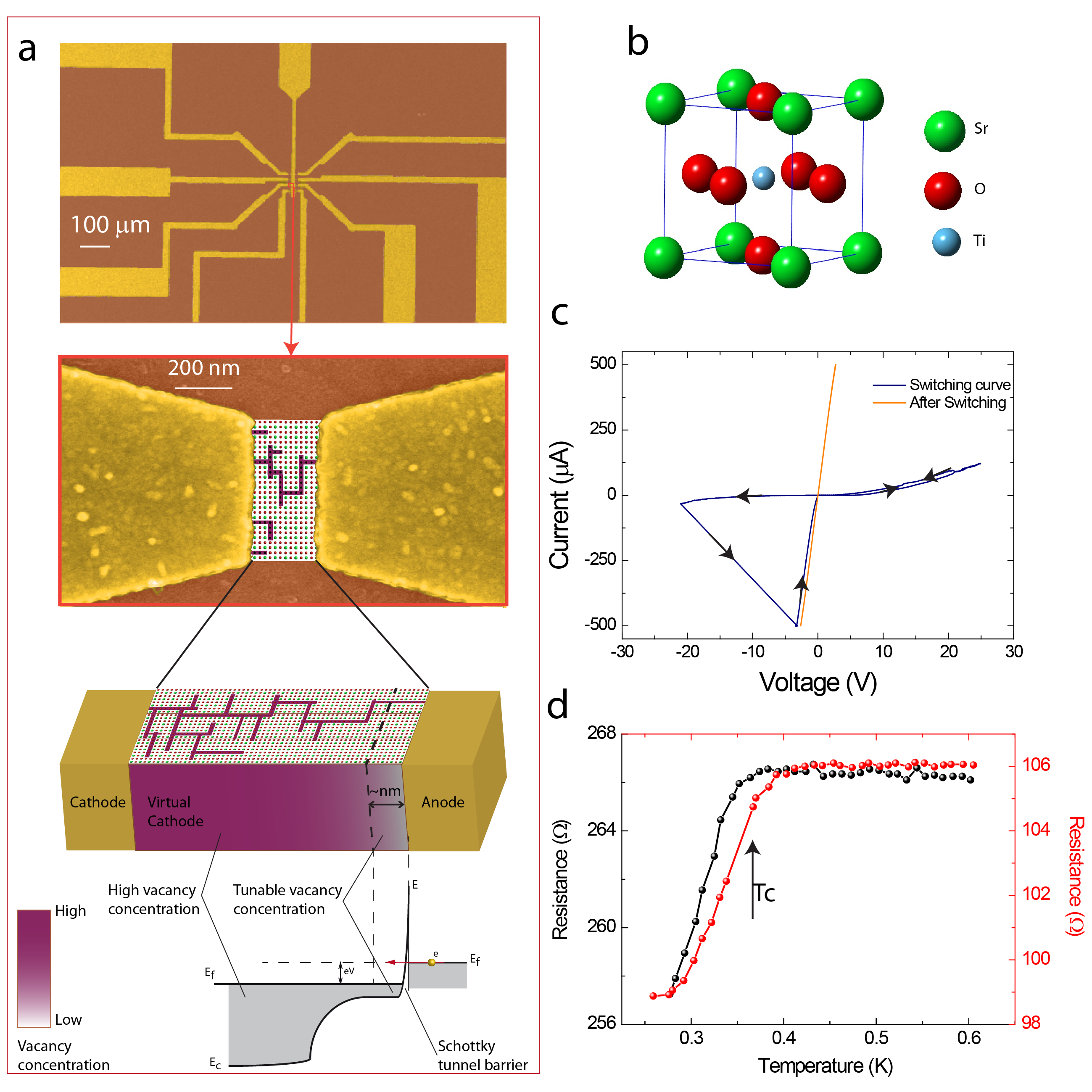}
\caption{\label{Fig1} (Color online) a) False color SEM image of gold electrodes deposited on the STO substrate. Applying a voltage between the electrodes leads to the formation of conducting filament which is identified by a switching event in the IV curve shown panel c). b)  Perovskite structure of STO crystal. d) Two-wires resistance of two different conducting filaments measured at low temperature.
}
\end{center}
\end{figure}

At low temperature ($T\sim 260~mK$), the conductance of this junction decreases by 5 orders of magnitude upon applying a single positive voltage pulse, of magnitude 5 V and duration 1 ms, Fig. 2a. In figure 2, each differential conductance   $G(V)=dI/dV(V)$ shown is measured at small voltage bias just after a voltage pulse has been applied. Up to 1600 pulses followed by measurements of the differential conductance have been realized for this junction.

As the junction is now insulating, applying negative voltage pulses moves the vacancies in the opposite direction and leads now to an increase of the conductance, up to 5 orders of magnitude, Fig. 2. Upon reaching a high differential conductance, $G(0)\sim 2~mS$, a single positive voltage pulse ($\sim 5~V$) is applied again, which switches the junction insulating once more.  This protocol can be executed many times -- up to 10 times for this junction. For each run, the junction is tuned from insulating to the conducting state by applying negative voltage pulses of small magnitude. Between each run, a positive pulse is applied to turn the junction insulating.

\begin{figure*}[h!]
\begin{center}
\includegraphics[width=18cm,keepaspectratio]{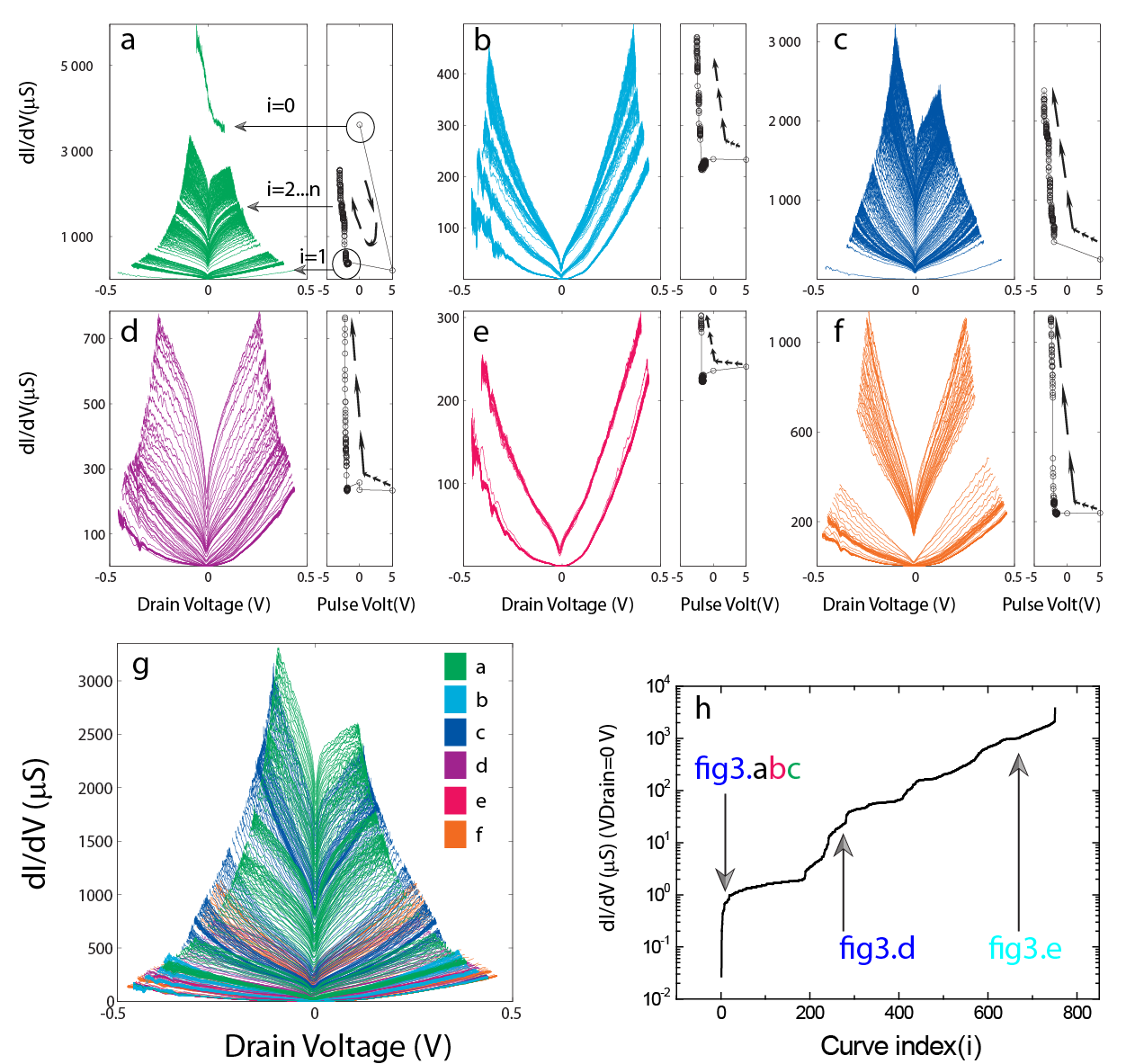}
\caption{\label{Fig2} (Color online) Panels a) to f) 750 differential conductance curves, indexed by i=0...n (n=749), measured at low temperature $T\sim 260~mK$, are shown as function of drain voltage. Before measurement of each curve, a voltage pulse of 1 ms is applied. For each curve, the amplitude of the pulse is indicated on the panel to the right of each differential curve panel. Panel a) The first differential curve measured (i=0) shows that the junction is highly conducting. A voltage pulse of 5 V turns the junction insulating, as shown by the curve (i=1). Then, applying many negative voltage pulses, from -2 V to -3 V, turns the junction conducting again as shown by the increasing value of the differential conductance between each curve. Between panel a) curves and panel b) curves, one positive voltage pulse has been applied which turned the junction insulating again. Applying negative voltage pulses led again to a progressive increase in the differential conductance. This protocol has been repeated six times, panels a) to f). The transition from the insulating state to the conducting state can be reproducibly repeated as shown by panel g) which shows that all the differential conductance curves, i=0...n (n=749), plotted separately in panels a) to f), can be superimposed on each other. Panel h) shows that the zero bias conductance, measured at $V_{Drain}=0~V$, increases by 5 orders of magnitude from the insulating to the conducting sample. Arrows labelled, fig3abc, fig3.d and fig3.e, indicate the zero bias conductance for the junctions whose DoS in shown in fig3.
}
\end{center}
\end{figure*}

We argue now that the effect of the voltage pulses is to migrate the oxygen vacancies only in a narrow region in close proximity to the electrode. First, this experiment is performed at low temperature where thermal activated migration of ions is completely inhibited and vacancy motion is due to electric-field activated migration of ions. As electric-field induced migration of ions can only occur on a length of the order of the Debye screening length, and because the screening length is short in conducting filaments, ions migration only occurs on a short length scale. This leads to a device with characteristics depicted Fig. 1a, where a nanosized region of weakly doped STO is located between two conducting electrodes. This  picture is consistent with previous explanations of memristive behaviors in titanium oxide based devices where the resistance switching is believed to occur in the immediate vicinity of the electrodes. The weakly doped STO nanosized region is connected on one side by an ohmic contact to a highly doped-STO electrode, and, on the other side, is connected by a Schottky barrier to the gold electrode, where the Schottky barrier makes a tunnel barrier for electrons~\cite{Wolf2011}. At low doping, the barrier height is given by the energy difference between the works functions of STO $\phi_{STO}=4.2~eV$, ref.~\cite{zagonel2009} and gold $\phi_{gold}=5.2~eV$, ref.~\cite{Huber1966}, which gives $\phi \sim 1~eV$. As we will see below, features of interest in the DoS are located at low energy, $\varepsilon < 0.1~eV$, implying that the IV curves can be interpreted as DoS curves. As the doping concentration increases in this n-doped semiconductor, the barrier height increases, while its width decrease~\cite{sze2007}, and so the junction remains in the tunneling regime. In this geometry, the differential conductance curves shown Fig. 2 represent, at low voltage, $V<100~meV$, the DoS of the weakly doped region:

\begin{equation}\label{Eq1}
\frac{G(V)}{G_0}=\int{\frac{N(\varepsilon)}{N_0}\frac{\partial f(\varepsilon-eV,T)}{\partial eV}d\varepsilon}
\end{equation}

%Thus, this method allows following precisely the evolution of the DoS in the nanosized region of STO located in the immediate vicinity of the gold electrode as the oxygen vacancies' concentration is tuned by applying voltage pulses.

When a voltage of positive polarity is applied, oxygen vacancies are repelled at the junction and the DoS decreases, for a voltage pulse of negative polarity, oxygen vacancies accumulate at the junction and the DoS increases.
Within a single run, sudden jumps of conductance are occasionally observed, as observed Fig. 2f. They correspond to resistance switching usually observed in memristive devices. These jumps do not appear to be correlated with any specific values of the conductance. As the 6 different runs shown in Fig. 2 are plotted together, Fig. 2g, we can see, first: that similar conductance curves can be reproduced even after a full cycle from insulating to conducting, which indicates that doping can be tuned reversibly in those junctions; second, that the junction conductance can be tuned continuously to any value as shown by the plot of conductance at zero bias for all measured conductance curves, Fig. 2h. A detailed look at the differential conductance curves taken at different conductance values allows to clearly characterize the evolution of the DoS in STO as a function of doping, Fig. 3.

\begin{figure}[h!]
\begin{center}
\includegraphics[width=7cm,keepaspectratio]{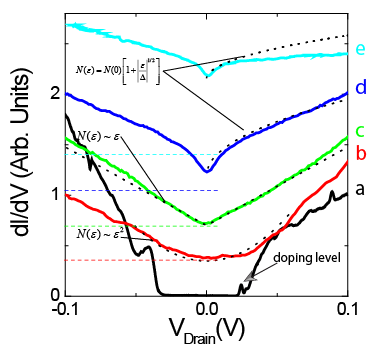}
\caption{\label{Fig3} (Color online) DoS at low bias voltage ($-0.1~V$ to $0.1~V$) at various steps of the transition from the insulator to the metal. The curves have been normalized at $V_{Drain}=0.1~V$ and displaced for clarity. The horizontal dashed lines indicate the level of zero conductance. Curve labelled a) is measured at very low doping concentration; it shows a hard gap and discrete electronic levels. At high carrier concentration, curve b), the DoS has a quadratic energy dependence, $N(\varepsilon) \sim \varepsilon^2$ , (Efros and Shklovskii) , as shown by the dotted line. At very high carrier concentration, curves d) and e), the DoS becomes finite at zero energy  with a cusp that can be fitted by Altshuler and Aronov law $N(\varepsilon)=N(0)[1+\left|\frac{\varepsilon}{E_T}\right|^{1/2}]$, as shown by the dotted lines. $E_T \sim 0.1~eV$ for curve e) $E_T \sim 0.003~eV$ for curve d). Finally, just at the critical point of the MIT, the DoS has a linear dependence with energy, $N(\varepsilon) \sim \varepsilon$ , dotted line.
}
\end{center}
\end{figure}

At the lowest doping, $G(V)$ shows a sharp gap and well resolved peaks at voltage values ($-41~mV$, $24~mV$, $78~mV$). These peaks, which have been observed in a second sample (Fig. S1), correspond to discrete doping levels. Applying a magnetic field leads to a Zeeman splitting of the peaks, Fig. 4, which can be fitted approximately by the relation  $\delta\varepsilon=m_{\ell}\mu_BB+2m_sB$, using for the orbital quantum numbers $m_{\ell}=-2,+1,-1$ and for the spin quantum numbers $m_s=\pm1/2$. As the orbital number   $m_{\ell}=-2$ has to be included to fit the Zeeman splitting, this indicates that this discrete doping level arises from doped electronic  states  into the $t_{2g}$ conduction band composed of the $3d_{xy}$, $3d_{xz}$ and $3d_{yz}$ titanium orbitals~\cite{Mattheiss1972}. This is further confirmed by the observation that this state is located just below the gap edge as expected for n-type dopants. Only samples annealed at high temperature\cite{lee1975,spinelli2010} have hydrogenic donor levels located very close to the conduction band. In non-annealed samples, as in our case, from an analysis of the temperature dependence of the Hall coefficient\cite{Tufte1967,lee1975} and the observation of sharp peaks in optical absorption spectrum\cite{lee1975}, it has been concluded that the oxygen vacancies levels may be located almost anywhere from nearly zero to about $0.1~eV$ below the conduction band, as observed in our tunneling spectrum.

\begin{figure}[h!]
\begin{center}
\includegraphics[width=8cm,keepaspectratio]{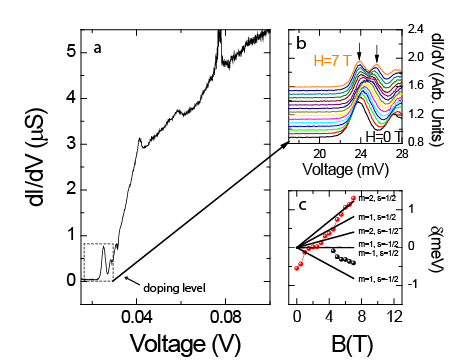}
\caption{\label{Fig4} (Color online) a) DoS at low bias voltage ($18~mV$-$100~mV$) in the low doping regime showing a sharp discrete doping level. b): Zoom ($18~mV-28~mV$) on the discrete electronic level. The peak splits in two peaks due to Zeeman splitting. The magnitude of the Zeeman splitting suggests that these electronic levels arises from a d ($\ell=2$) orbital level, shown panel c).
}
\end{center}
\end{figure}

While the observation of individual dopant states was attempted through the formation of nano-sized Schottky diodes~\cite{Smit2004,Calvet2002} and nanosized field effect transistors~\cite{Calvet2011,Calvet2007}, we find that the use of oxide semiconductors, with its possibility of fine-tuning the dopant concentration in-situ at low temperature, offers a remarkable tool for the study of individual dopant states.
We show now this method allows studying the MIT as a function of doping with unprecedented resolution. As the doping concentration increases, the gap fills in and sharp features disappear, however, a soft gap remains where the DoS vanishes at the Fermi level with the quadratic form $N(\varepsilon) \sim \varepsilon^2$ in 3D, due to Coulomb interactions as shown by Efros and Shklovskii~\cite{Efros1975}. Such a soft gap was observed previously in doped semiconductors through tunneling spectroscopy~\cite{Lee2004a,Bielejec2001,Lee1999a,Massey1996,McMillan1981}.  As the dopant concentration increases, the soft gap closes and a transition to a metallic regime occurs. In the metallic regime, the DoS is finite at the Fermi level and a cusp, $N(\varepsilon)=N(0)[1+\left|\frac{\varepsilon}{E_T}\right|^{1/2}]$ in 3D, is observed, where the correlation energy parameter $E_T=\hbar D/\ell^2$, i.e. the Thouless energy, depends on the diffusion constant $D$ and the electron mean free path $\ell$. This cusp is due to the exchange part of the Coulomb interactions as shown by McMillan~\cite{McMillan1981a}, Altshuler and Aronov~\cite{Altshuler1979}. Fitting the data shows that $E_T$ decreases at the approach of the transition, as expected when the electronic states at the Fermi level gets more and more localized. The maximum value found for $E_T\sim 0.1~eV $ can be explained assuming the reasonable values for $D\sim 2~cm^2~s^{-1}$ and $\ell\sim1~nm$. Finally, just at the transition, the critical DoS follows a linear behaviour, $N(\varepsilon) \sim \varepsilon$ . The question of the energy dependence of the critical DoS remains one of the major theoretical question of the Mott-Anderson MIT~\cite{Belitz1994}. A theoretical prediction suggests~\cite{Lee1999a} that the DoS should follow $N_c(\varepsilon) \sim \varepsilon^{(3/\eta)-1}$; this implies that $\eta \sim 1.5$, which is within the bounds 1$<\eta<$3 , as theoretically expected~\cite{McMillan1981}. This power law suppression of the DoS at criticality is a signature of multifractality, a remarkable property of electronic wavefunctions near the MIT\cite{Evers2008}. Recent theoretical works have found that multifractality persists even in presence of Coulomb interactions\cite{Burmistrov2013,Amini2013} and multi-fractal characteristics have been found in STM spectroscopy measurements of Ga$_{1-x}$Mn$_x$As\cite{Richardella2010}.

Note that the DoS curves were compared with theoretical predictions assuming tri-dimensionality. Indeed, while previous works\cite{Waser2007,Yang2013} have shown that the vacancies are distributed along filamentary paths, we don't expect that quantum confinement effects would alter the 3D band structure of STO along the filament. For this reason, we can assume that the effect of doping is to inject carriers into the 3D band structure of STO.

While we don't have a direct measure of the oxygen vacancy concentration, we can estimate that this concentration changes from nearly zero, where single impurity levels are observed, to $10 ^{20}~e~cm^{-3}$, where the superconducting transition temperature $T_c\sim 350~mK$ is maximum\cite{Ueno2011}. The MIT should be at concentrations below $n \sim 10^{16}~e~cm^{-3}$ according to results from Spinelli et al. \cite{spinelli2010}.

%This method should apply to a variety of other oxides, including materials presenting strong electron correlations such as the cuprates or the manganites.

%To resume, we have found that oxygen vacancies could be migrated by an electric field at very low temperature in nanosized metal-oxide-metal junctions. As the undoped region of the junction is connected by a Schottky tunnel barrier to the metal electrode, the current-voltage characteristics provides the DoS. Upon applying voltage pulses, we find possible to tune continuously the DoS across the MIT where all the characteristics expected in the vicinity of the MIT are found: the soft Coulomb gap on the insulating side and the Altshuler - Aronov corrections on the metallic side.

We thank M. Rosticher and J. Palomo for their help with clean room works.
We acknowledge support from ANR grant "QUANTICON" 10-0409-01 and ANR Grant "CAMELEON" 09-BLAN-0388-01.

\bibliography{STOpaper}

\end{document}